\documentclass{article}
\usepackage{spconf,amsmath,graphicx}
\usepackage{amssymb,amsfonts}
\usepackage{algorithmic}
\usepackage{graphicx}
\usepackage{textcomp}
\usepackage{xcolor}
\usepackage{amssymb}
\usepackage{amsmath}
\usepackage{float}
\usepackage{subfig}
\usepackage{url}
\usepackage{hanging}
\usepackage{makecell}

\title{Quality analysis of the CODING BITRATE TRADEOFF BETWEEN GEOMETRY AND ATTRIBUTES FOR COLORED POINT CLOUDS}%
\name{Joao Prazeres, Rafael Rodrigues, Manuela Pereira, Antonio M. G. Pinheiro\thanks{Research funded by the Portuguese FCT-Fundacao para a Ciencia e Tecnologia  under the project UIDB/50008/2020, PLive  X-0017-LX-20.}}
\address{Universidade da Beira Interior \& Instituto de Telecomunicacoes, Covilha, Portugal}
\begin{document}
%
\maketitle
\begin{abstract}
Typically, point cloud encoders allocate a similar bitrate 
for geometry and attributes (usually RGB color components) information coding.
This paper reports a quality study considering different coding  bitrate tradeoff between geometry and attributes.
%
A set of five point clouds, representing different characteristics and types of content 
was encoded with the MPEG standard Geometry Point Cloud Compression (G-PCC), using octree to encode geometry information, and both the Region Adaptive Hierarchical Transform and the Prediction Lifting transform for attributes. Furthermore, the JPEG Pleno Point Cloud Verification Model was also tested.
Five different attributes/geometry bitrate tradeoffs were considered, notably 70\%/30\%, 60\%/40\%, 50\%/50\%, 40\%/60\%, 30\%/70\%.
Three point cloud objective metrics were selected to assess the quality of the reconstructed point clouds, notably the PSNR YUV, the Point Cloud Quality Metric, and GraphSIM.
Furthermore, for each encoder, the Bjonteegaard Deltas were computed for each tradeoff, using the 50\%/50\% tradeoff as a reference. The reported results indicate that using a higher bitrate allocation for attribute encoding usually yields slightly better results.
\end{abstract}
\section{Introduction} \label{sec:intro}

Digital imaging technologies have expanded the ability to capture 3D scenes by combining photogrammetry and computer graphics. Colored point clouds are widely used for accurate representation of these scenes, regardless of viewpoint and interaction. These clouds consist of geometry coordinates and color attributes, and require efficient coding solutions and quality models for accurate benchmarking. 

However, there has been limited research conducted to assess different compression tradeoffs between geometry and color attribute bitrate when coding colored point clouds. Typically, the most established point cloud codecs, namely MPEG G-PCC and V-PCC, use a 50\% tradeoff~\cite{VPCCandGPCC}.
Previous subjective quality evaluation studies \cite{icip2020,PrazeresACM2022,BASICS,PRAZERESSPIC2024} seems to indicate that static colored point clouds benefit from high-quality color attribute reconstruction, providing a higher quality of experience~\cite{le2012qualinet} when visualizing a point cloud. 

\begin{figure}[t!]
    \centering
    \subfloat[\textit{Bouquet}]{\includegraphics[width=0.3\linewidth]{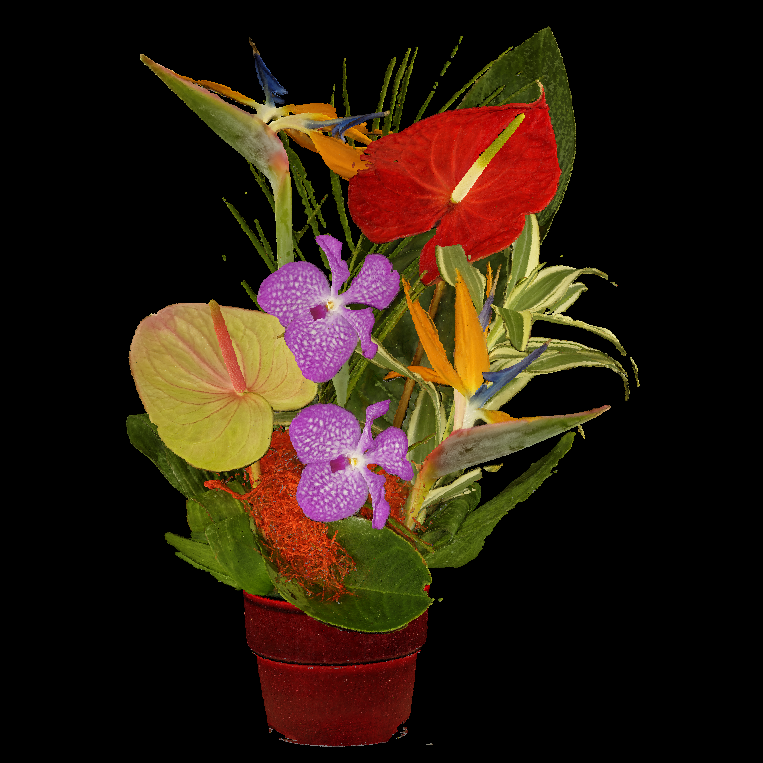}}\quad
    \subfloat[\textit{StMichael}]{\includegraphics[width=0.3\linewidth]{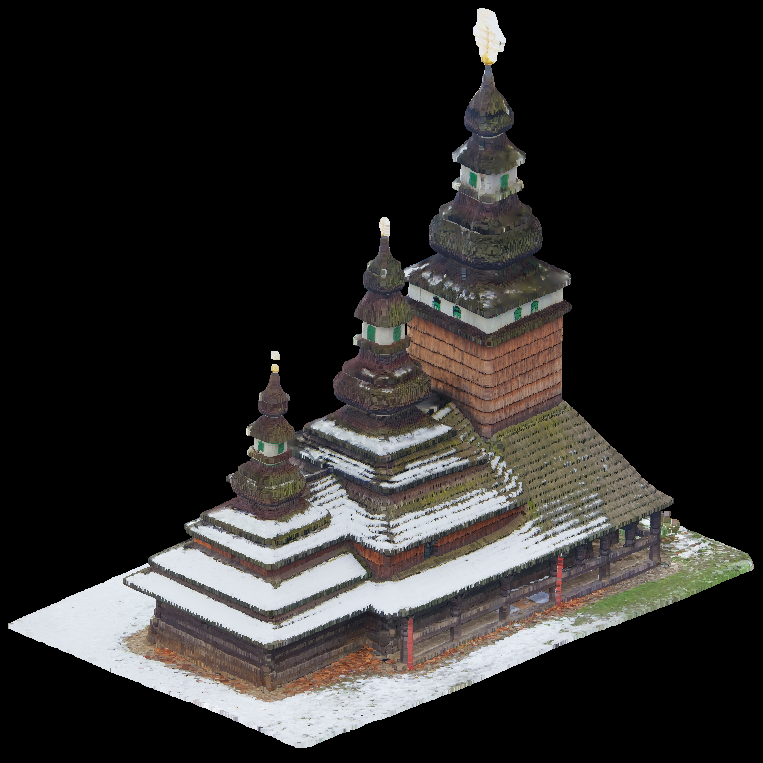}}\quad
    \subfloat[\textit{Thaidancer}]{\includegraphics[width=0.3\linewidth]{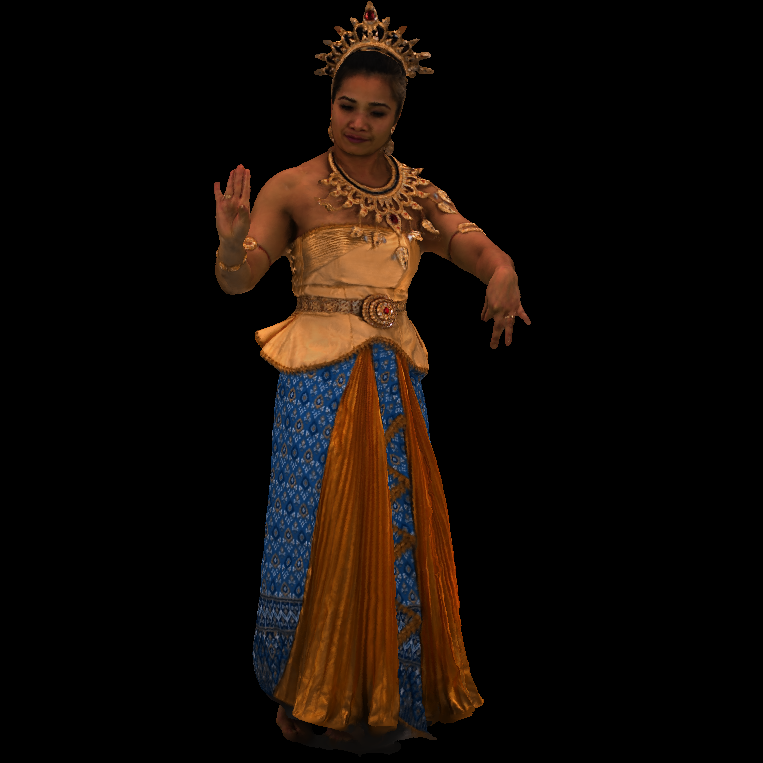}} \\
    \subfloat[\textit{CITIUSP}]{\includegraphics[width=0.3\linewidth]{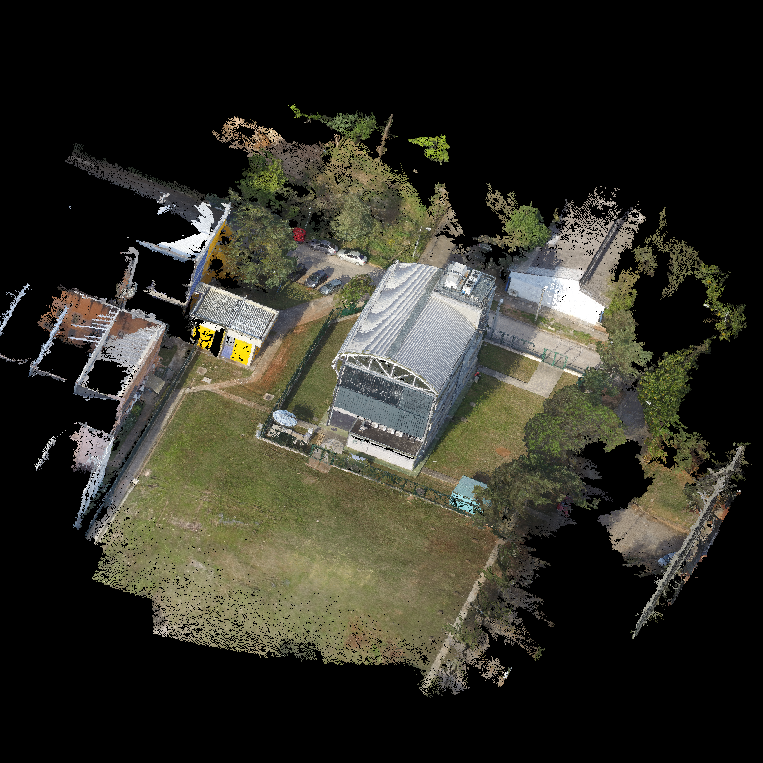}}\quad
    \subfloat[\textit{Bumbameuboi}]{\includegraphics[width=0.3\linewidth]{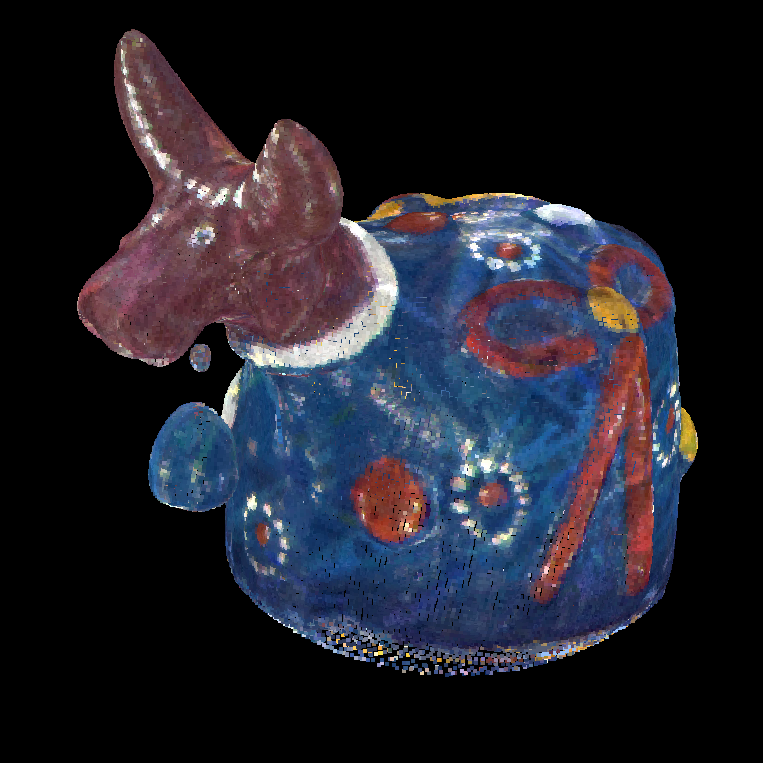}} \\
   \caption{Point Cloud testing set.}
    \label{PCS}
\end{figure} 

\begin{table}[t]
    \caption{Point cloud statistics}
    \huge
    \centering
    \resizebox{\linewidth}{!}{%
    \begin{tabular}{|c|c|c|c|c|c|}
    \hline
    Point Cloud &\makecell[c]{Sparsity \\ (K=20)} & \makecell[c]{Color Gamut \\ Volume} & Y Deviation & Cb Deviation & Cr Deviation  \\ \hline
       \textit{Bouquet} & 1.72 & 4.71\% & 0.151 & 0.069 & 0.081  \\ \hline
       \textit{StMichael} & 1.72 & 3.46\% & 0.204 & 0.027 & 0.026  \\ \hline
       \textit{Thaidancer} & 1.9 & 3.33\% & 0.101 & 0.068 & 0.065  \\ \hline
       \textit{CITIUSP} & 5.01 & 2.43\% & 0.159 & 0.041 & 0.023  \\ \hline
       \textit{Bumbameuboi} & 6.661 &2.58\% & 0.132 & 0.070 & 0.067 \\ \hline
    \end{tabular}%
    }
    \label{table:statistics}

\end{table}

The main contribution of this paper is to provide an insight on the best bitrate tradeoff between color attributes and geometry by objectively assessing five different tradeoffs of color attribute/geometry, notably 70\%/30\%, 60\%/40\%, 50\%/50\%, 40\%/60\%, and 30\%/70\%.
Five different point clouds, shown in Fig. \ref{PCS} were selected.
The MPEG Geometry Point Cloud Compression (G-PCC)~\cite{VPCCandGPCC} standard, using the octree mode for geometry and the Regional Adaptive Hierarchical Transform (RAHT))~\cite{Queiroz2016a} and the predicting/Lifting (predlift)~\cite{VPCCandGPCC} transform for color attribute encoding, was used in this work. Moreover, the learning-based solution JPEG Pleno Point Cloud Verification model~\cite{VM-CD}, was also included in this study.
V-PCC was not used because it was not possible to control the tradeoff between bitrates independently.

\begin{figure}[t]
    \begin{center}

     \subfloat[\textit{30\%/70\% r01}]{%
         \includegraphics[width=0.32\linewidth]{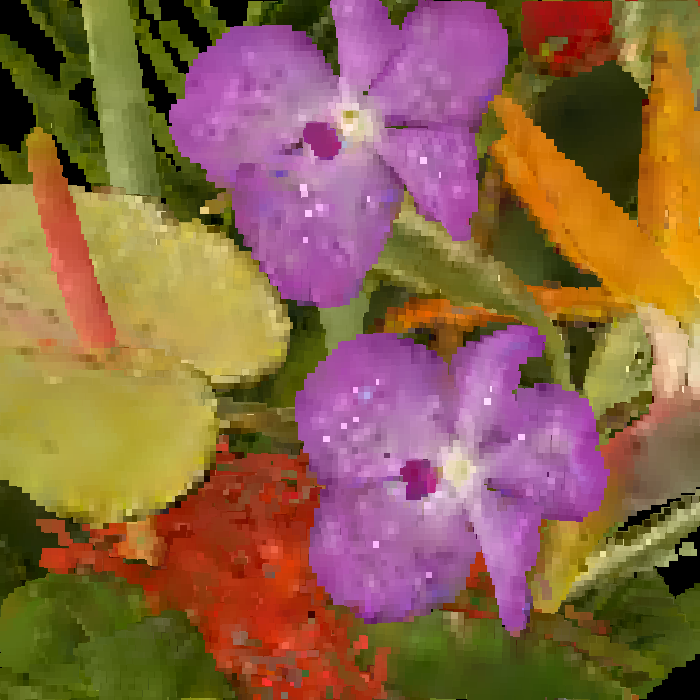}}
         \hfill
     \subfloat[\textit{30\%/70\% r03}]{%
         \includegraphics[width=0.32\linewidth]{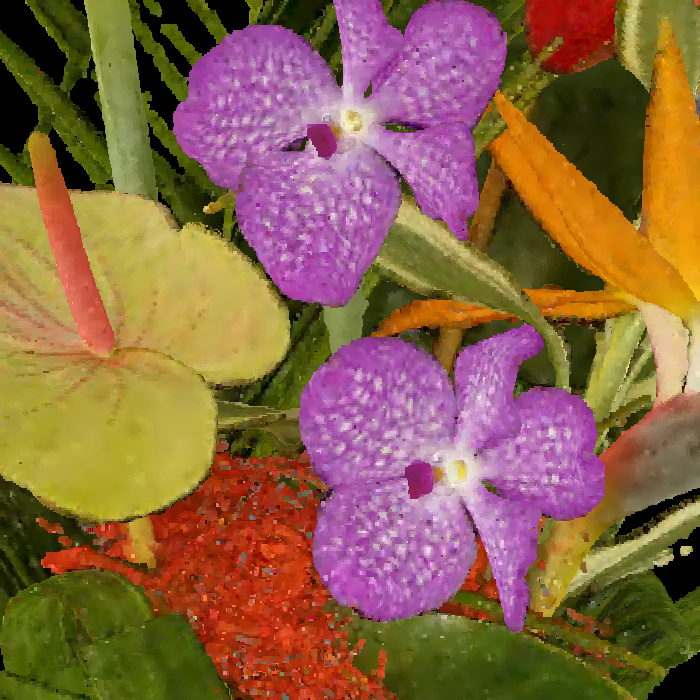}}
         \hfill
    \subfloat[\textit{30\%/70\% r04}]{%
         \includegraphics[width=0.32\linewidth]{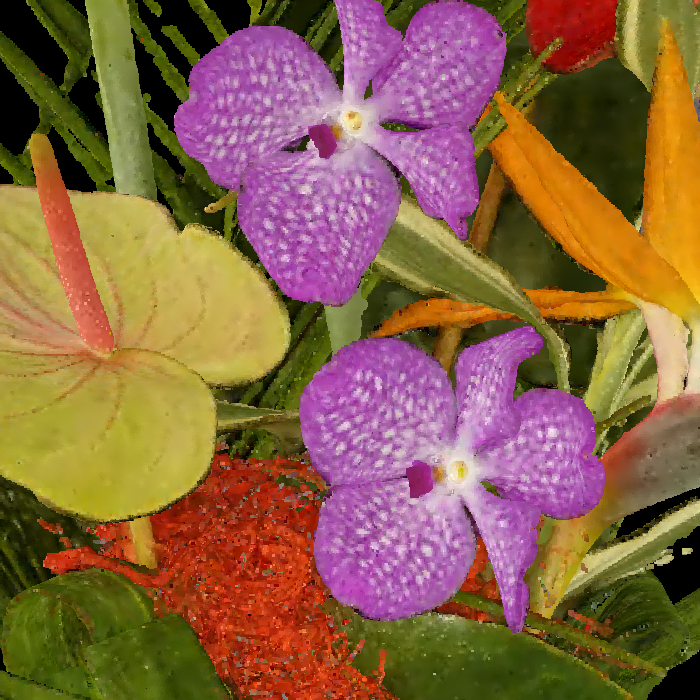}}
         \hfill \\
     \subfloat[\textit{50\%/50\% r01}]{%
         \includegraphics[width=0.32\linewidth]{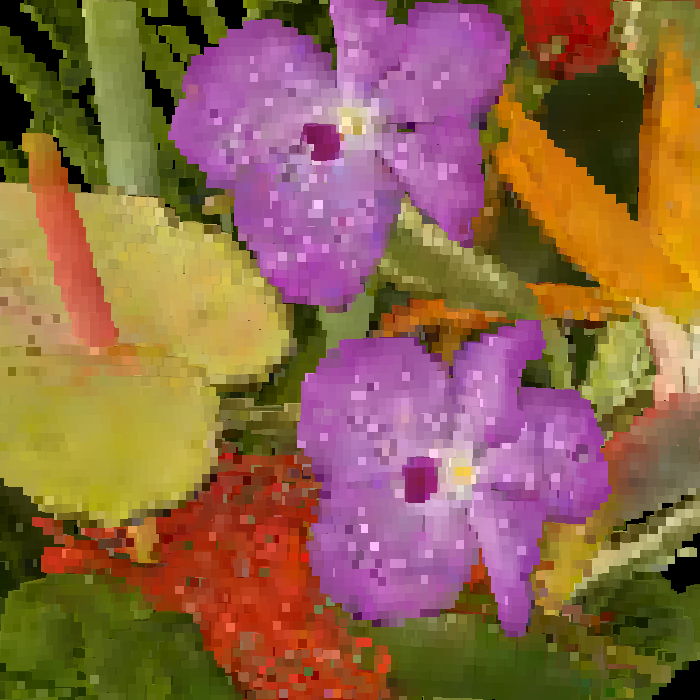}}
         \hfill
     \subfloat[\textit{50\%/50\% r03}]{%
         \includegraphics[width=0.32\linewidth]{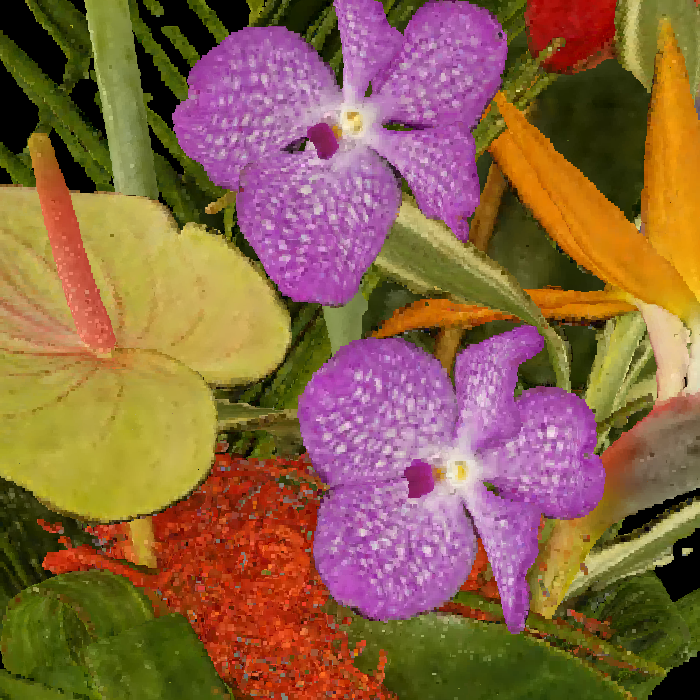}}
         \hfill
    \subfloat[\textit{50\%/50\% r04}]{%
         \includegraphics[width=0.32\linewidth]{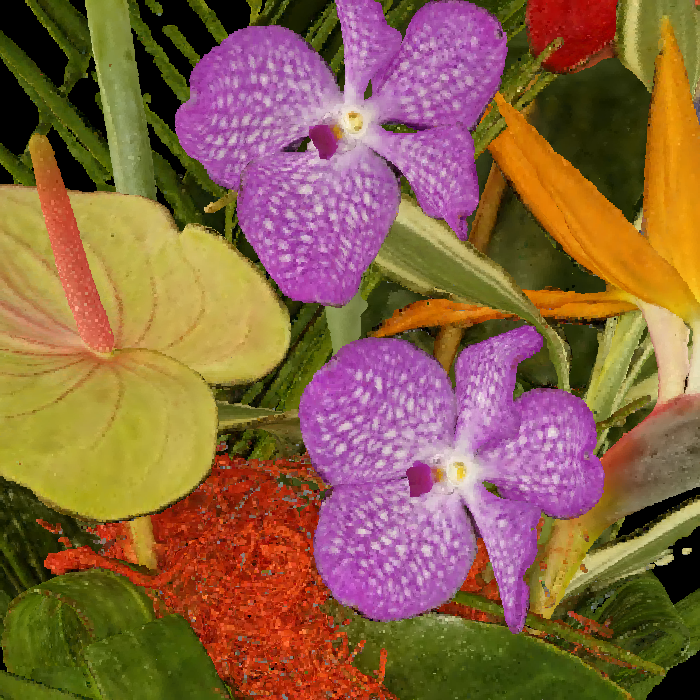}}
         \hfill \\
    \subfloat[\textit{70\%/30\% r01}]{%
         \includegraphics[width=0.32\linewidth]{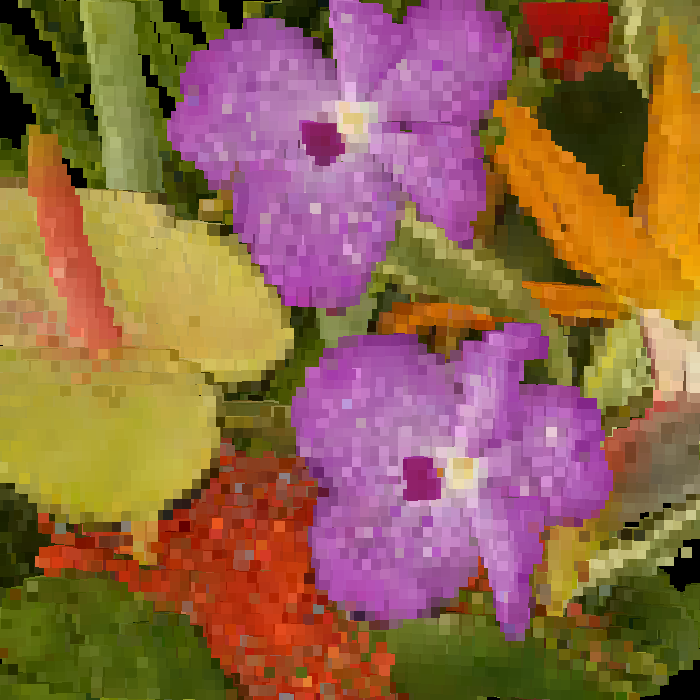}}
         \hfill
     \subfloat[\textit{70\%/30\% r03}]{%
         \includegraphics[width=0.32\linewidth]{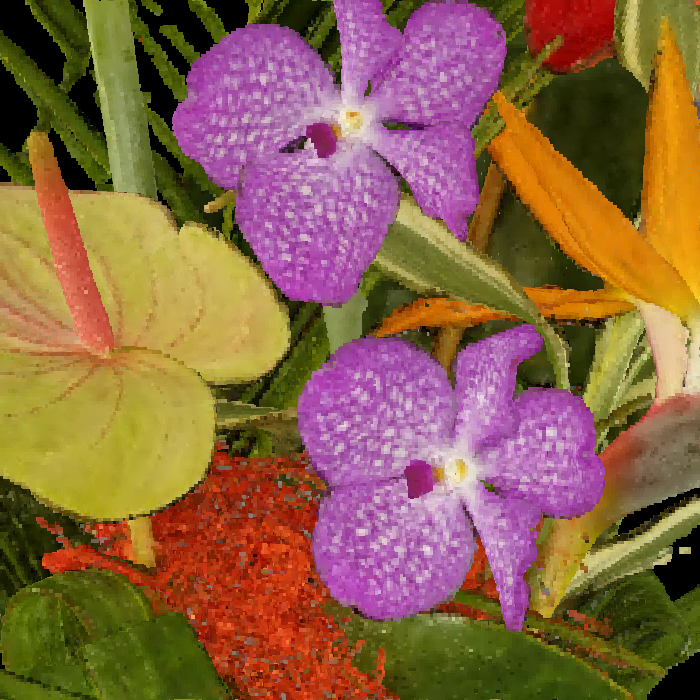}}
         \hfill
    \subfloat[\textit{70\%/30\% r04}]{%
         \includegraphics[width=0.32\linewidth]{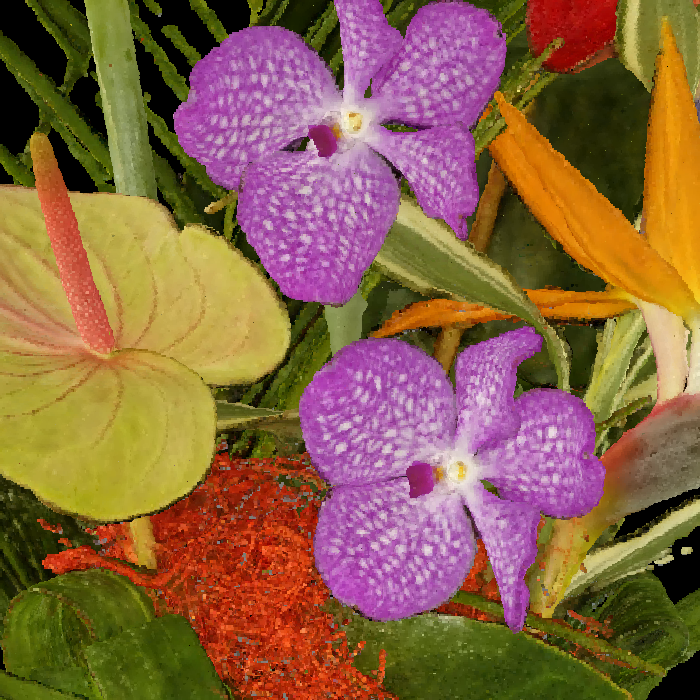}}
         \hfill \\
         \caption{Visual examples of the \textit{Bouquet} Point cloud encoded with G-PCC, using the predicting/Lifting transform, with some of the chosen attributes/geometry tradeoff.}\label{fig:VIS_pred}
    \end{center}
    
\end{figure}
\begin{figure}[htb]
    \begin{center}
        \subfloat[\textit{30\%/70\% r01}]{%
         \includegraphics[width=0.32\linewidth]{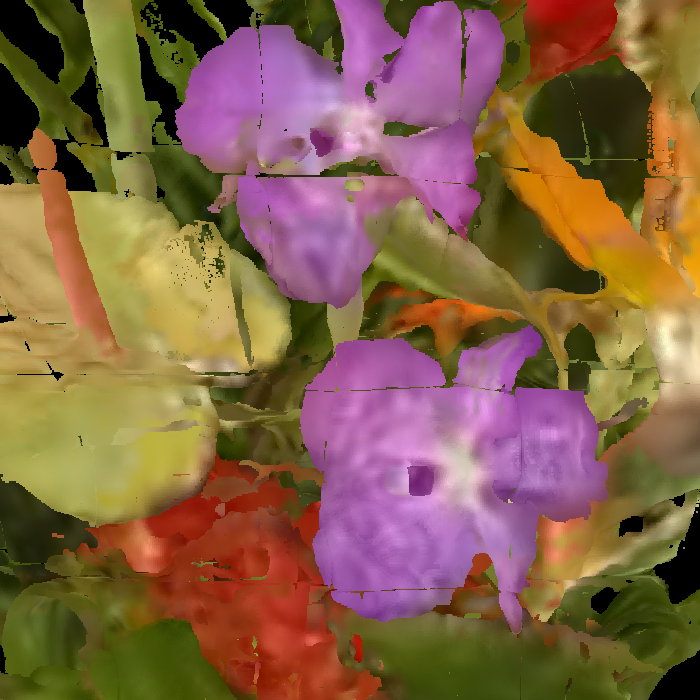}}
         \hfill
     \subfloat[\textit{30\%/70\% r03}]{%
         \includegraphics[width=0.32\linewidth]{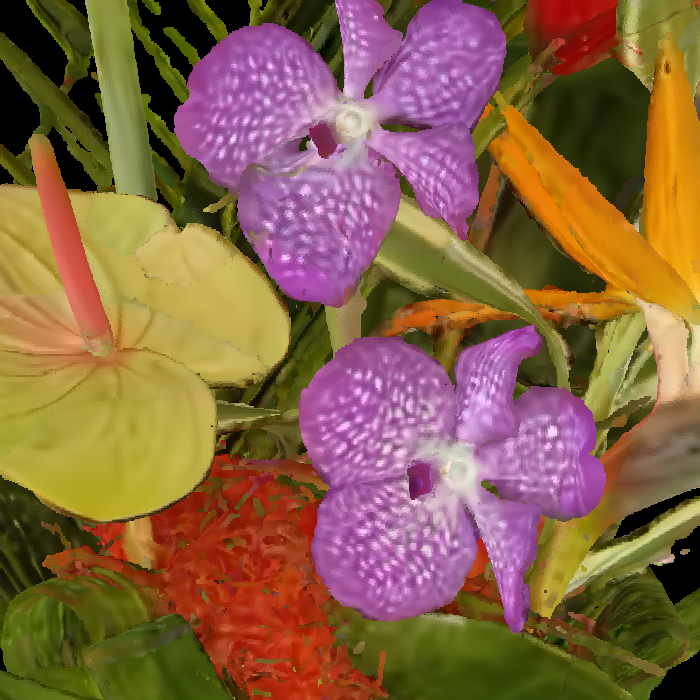}}
         \hfill
    \subfloat[\textit{30\%/70\% r03}]{%
         \includegraphics[width=0.32\linewidth]{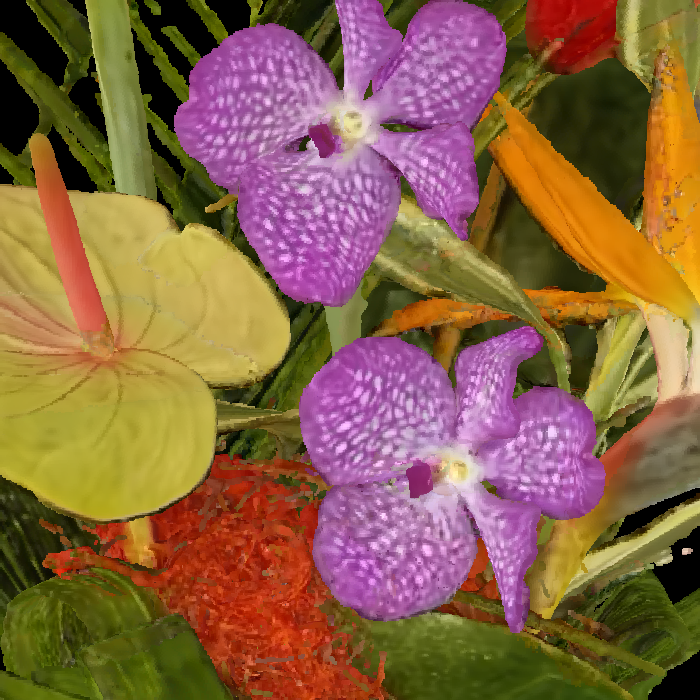}}
         \hfill \\
     \subfloat[\textit{50\%/50\% r01}]{%
         \includegraphics[width=0.32\linewidth]{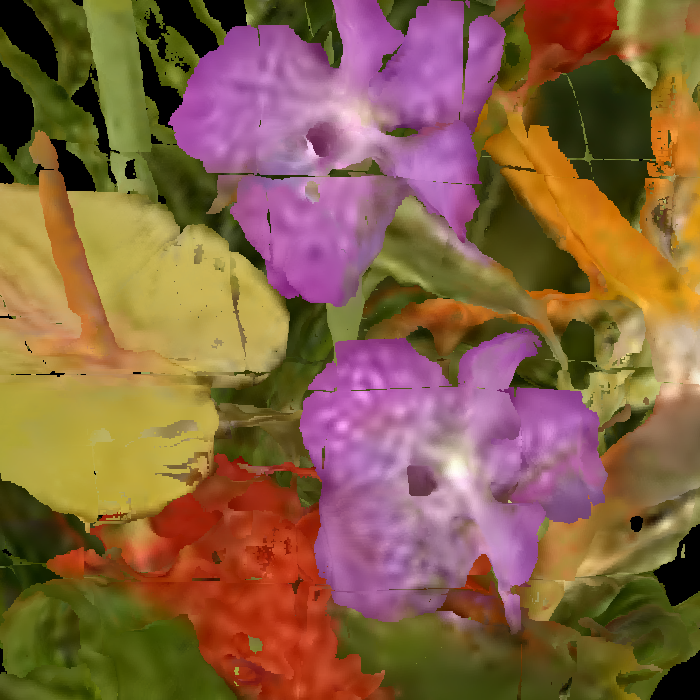}}
         \hfill
     \subfloat[\textit{50\%/50\% r03}]{%
         \includegraphics[width=0.32\linewidth]{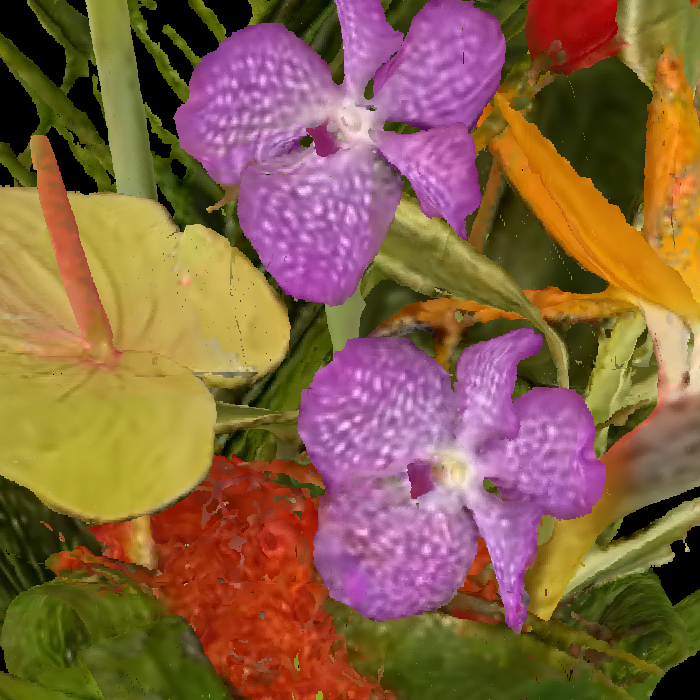}}
         \hfill
    \subfloat[\textit{50\%/50\% r04}]{%
         \includegraphics[width=0.32\linewidth]{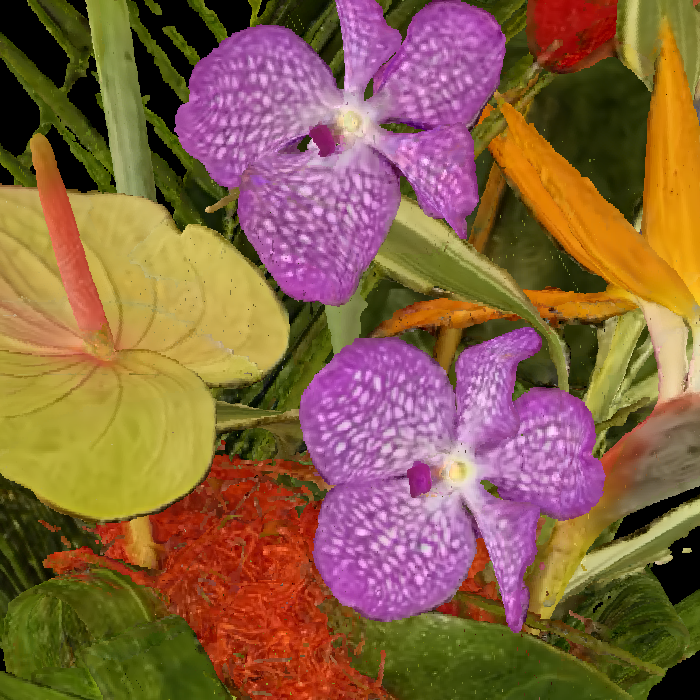}}
         \hfill \\
    \subfloat[\textit{70\%/30\% r01}]{%
         \includegraphics[width=0.32\linewidth]{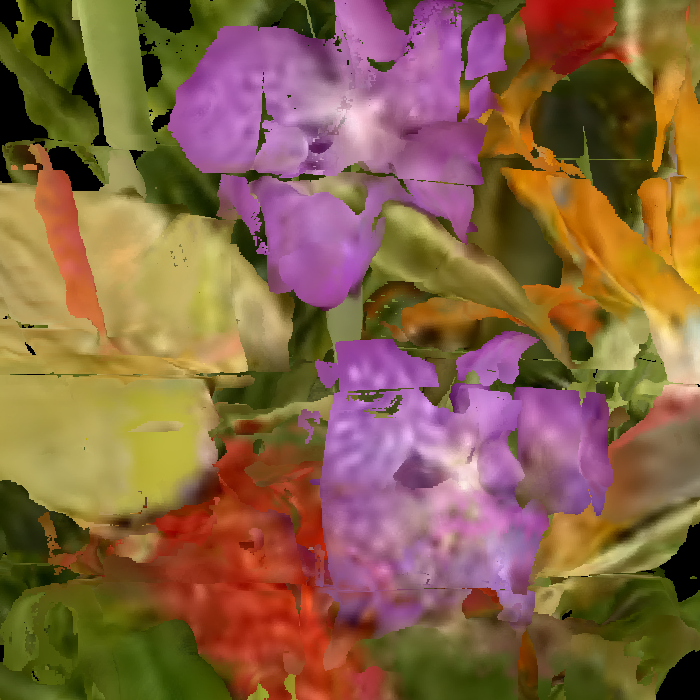}}
         \hfill
     \subfloat[\textit{70\%/30\% r03}]{%
         \includegraphics[width=0.32\linewidth]{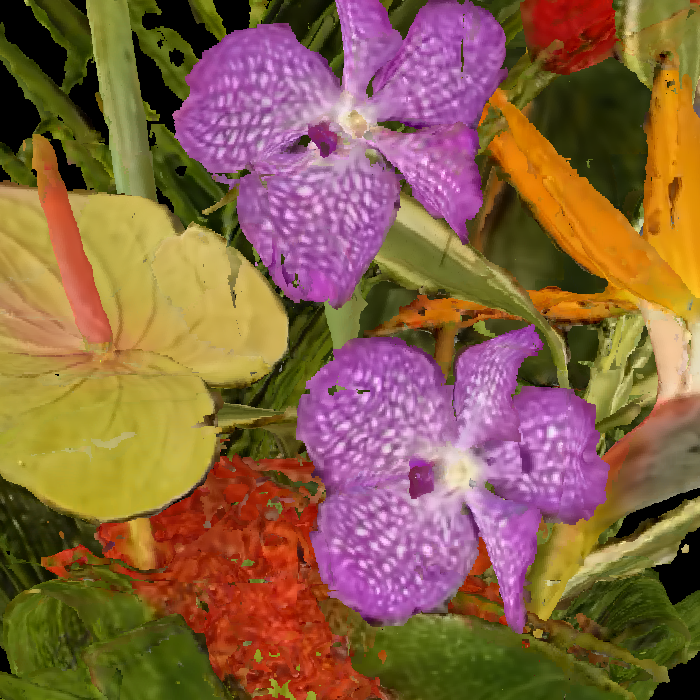}}
         \hfill
    \subfloat[\textit{70\%/30\% r04}]{%
         \includegraphics[width=0.32\linewidth]{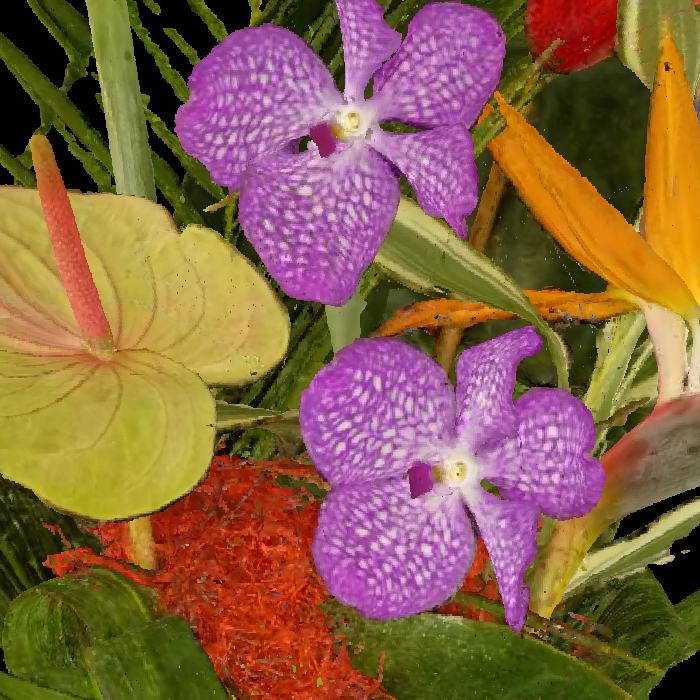}}
         \hfill \\
    \caption{Visual examples of the \textit{Bouquet} Point cloud encoded using the JPEG PCC VM, with some of the chosen attributes/geometry tradeoff.}\label{fig:VIS_VM}
    \end{center}
 \end{figure}

\section{Related work}
Numerous studies have been conducted to evaluate the quality of point clouds, taking into account several coding approaches and experimental configurations~\cite{LiuResSCNN2023,Su2023,projectionYANG,BASICS,lazzarotto2024subjective,WuIEEE}. Perry \textit{et al.} presented an assessment of the perceived quality of MPEG Point Cloud codecs, notably Video Point Cloud Compression (V-PCC) and Geometry Point Cloud Compression (G-PCC), using a 2D display~\cite{icip2020}. 
Recently, with the emerging of deep-learning technology, several solutions were developed~\cite{GuardaRS-DLPCC,quach2020improved,Jianqiang-PCGCv2} that target the encoding of point cloud geometry. 
They have been compared to G-PCC~\cite{PrazeresACM2022}, achieving a better performance.
Deep-learning technology has also been researched to encode point cloud attributes~\cite{DatNguyenTCSVT2023,WangMIPR2022}, but the solutions only achieve comparable performance to G-PCC. 
Recently, a joint geometry-color learning-based coding solution was proposed~\cite{GuardaTMM2023}. It reveals a competitive performance with G-PCC and V-PCC for geometry encoding, but it cannot surpass their performance in terms of attribute encoding.

\section{Evaluation Procedure}
\subsection{Dataset}
Fig. \ref{PCS} represents the set of five point clouds containing geometry and attribute information used in this study. The set consists of \textit{Bouquet} and \textit{StMichael} from ShapeNet~\cite{ShapeNet}, \textit{Thaidancer} from MPEG, and \textit{CITIUSP} and \textit{Bumbameuboi} from the University of São Paulo Point Cloud Dataset\footnote{http://uspaulopc.di.ubi.pt}. This set includes a diverse range of scenarios, depicting humans, objects, buildings, landscapes, and cultural heritage artifacts.

Table \ref{table:statistics} depicts several point cloud characteristics, notably the point cloud sparsity, color gamut volume, and standard deviation for each color channel of the YCbCr color space. The sparsity is defined as the average distance between each point and its 20 nearest neighbors, averaged over the entire point cloud.
The color gamut volume is defined as the volume of the convex hull of the distribution of color points in the YCbCr color space.
The test set reveals a good variation in both point cloud sparsity and color gamut variation.

\begin{figure}[t]
    \begin{center} 
    \subfloat[\textit{PSNR YUV (predlift)}]{\includegraphics[width=0.32\linewidth]{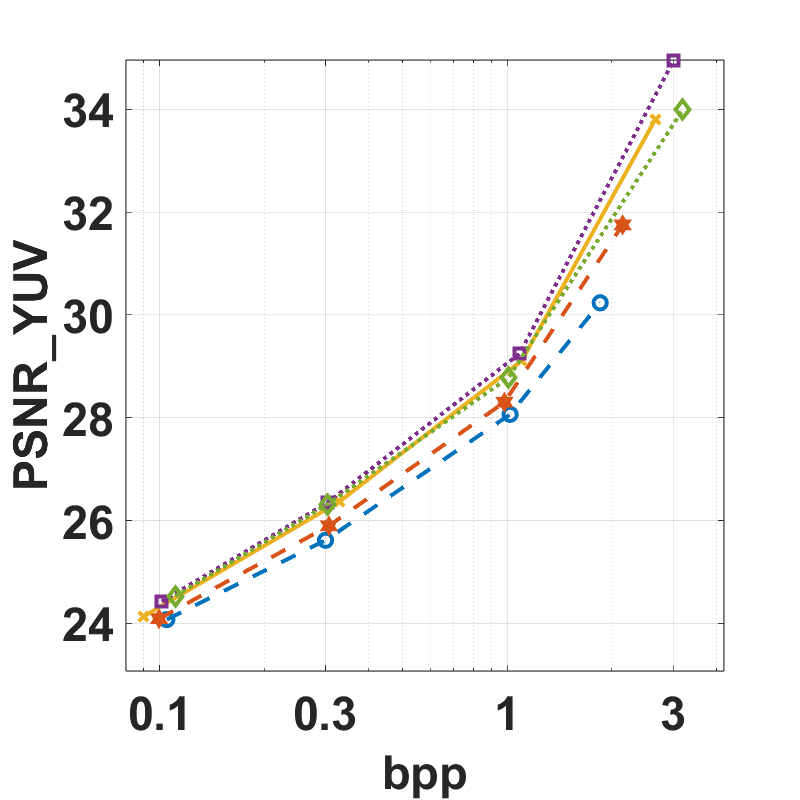}}\hfill
    \subfloat[\textit{GraphSIM (predlift)}]{\includegraphics[width=0.32\linewidth]{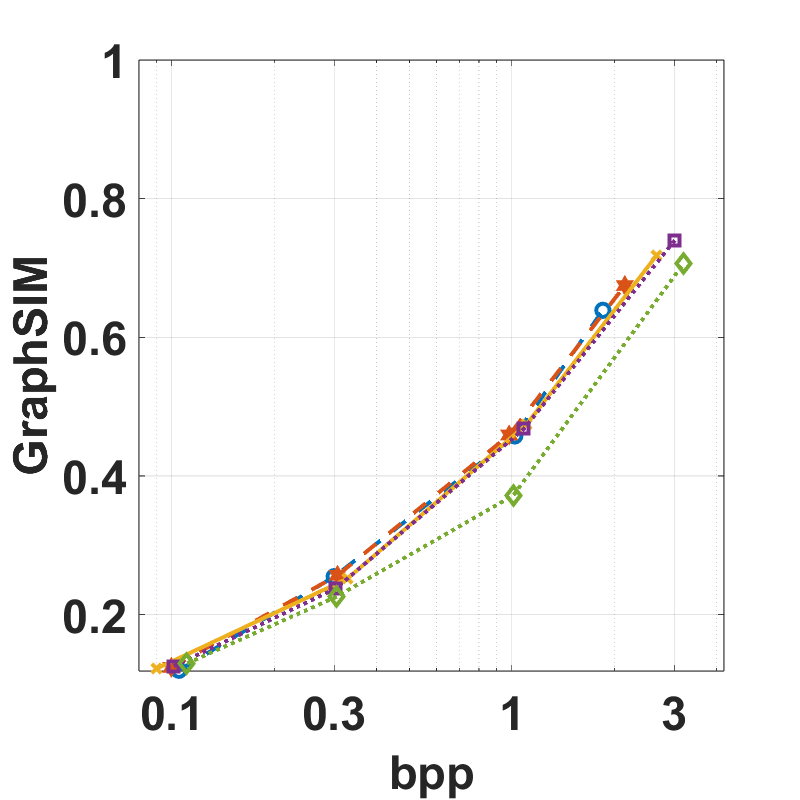}}\hfill
    \subfloat[\textit{1 - PCQM (predlift)}]{\includegraphics[width=0.32\linewidth]{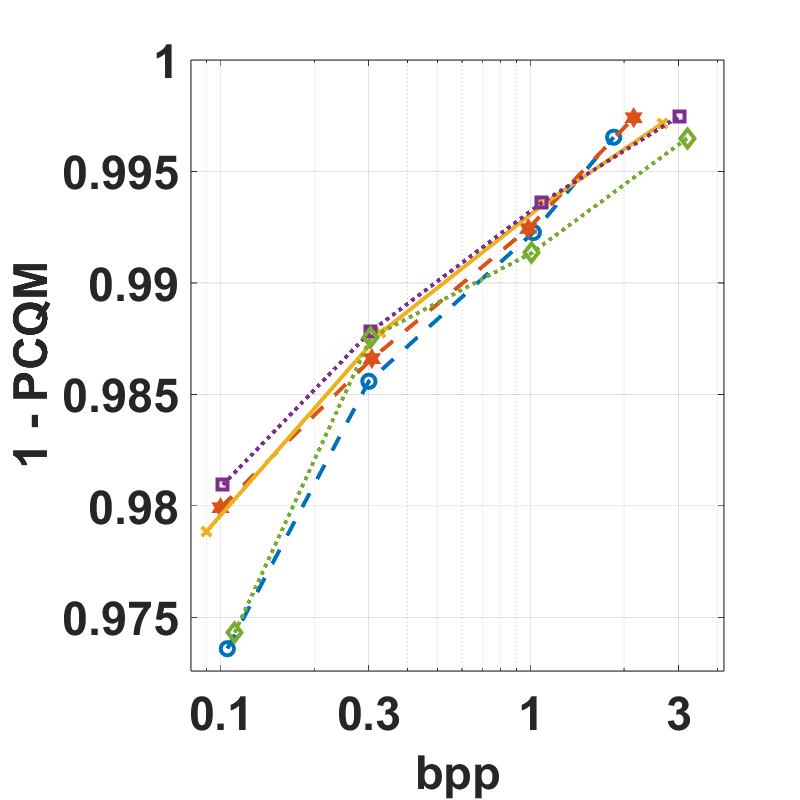}}\hfill \\
    \subfloat[\textit{PSNR YUV (RAHT)}]{\includegraphics[width=0.32\linewidth]{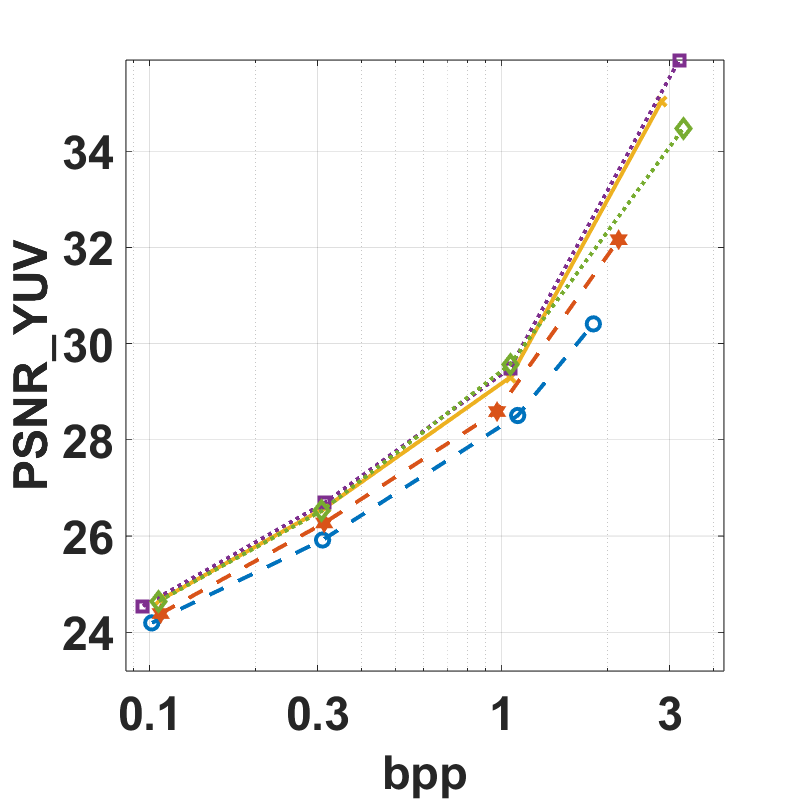}}\hfill 
    \subfloat[\textit{GraphSIM (RAHT)}]{\includegraphics[width=0.32\linewidth]{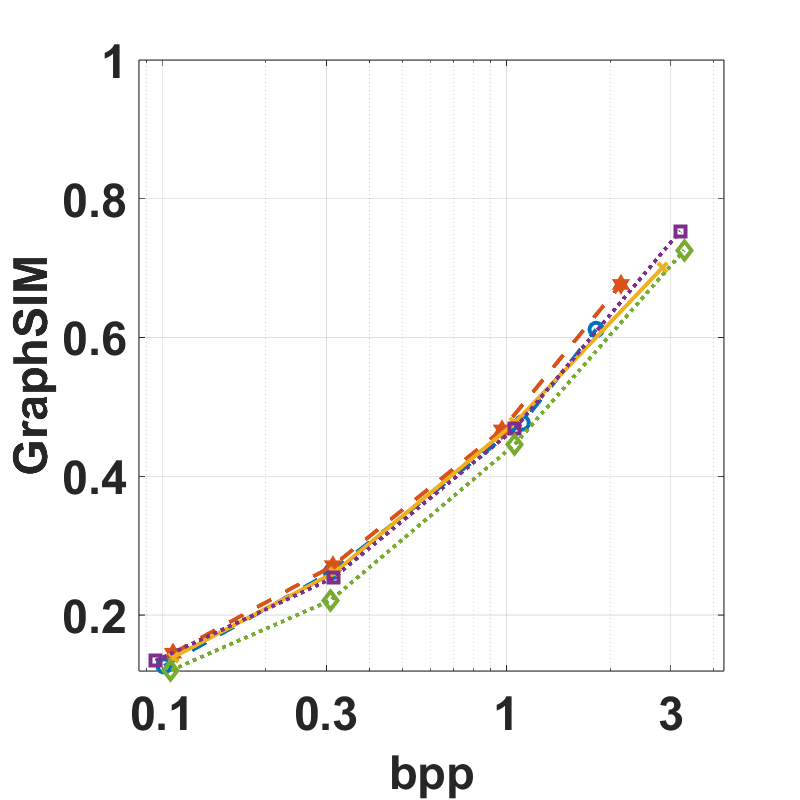}} \hfill
    \subfloat[\textit{1 - PCQM (RAHT)}]{\includegraphics[width=0.32\linewidth]{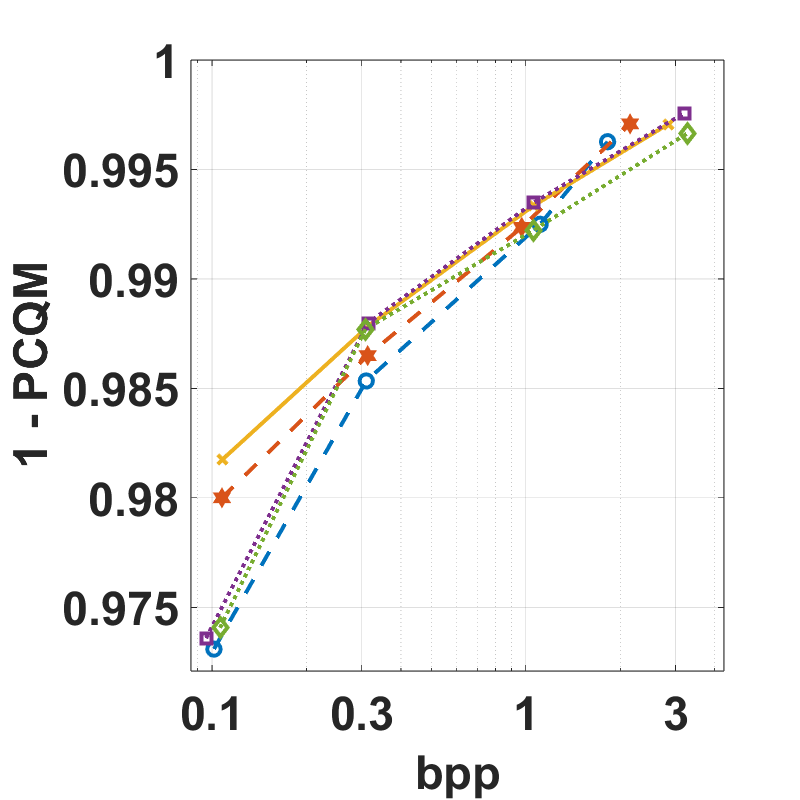}}\hfill\\
    \subfloat[\textit{PSNR YUV (JPEG PCC VM)}]{\includegraphics[width=0.32\linewidth]{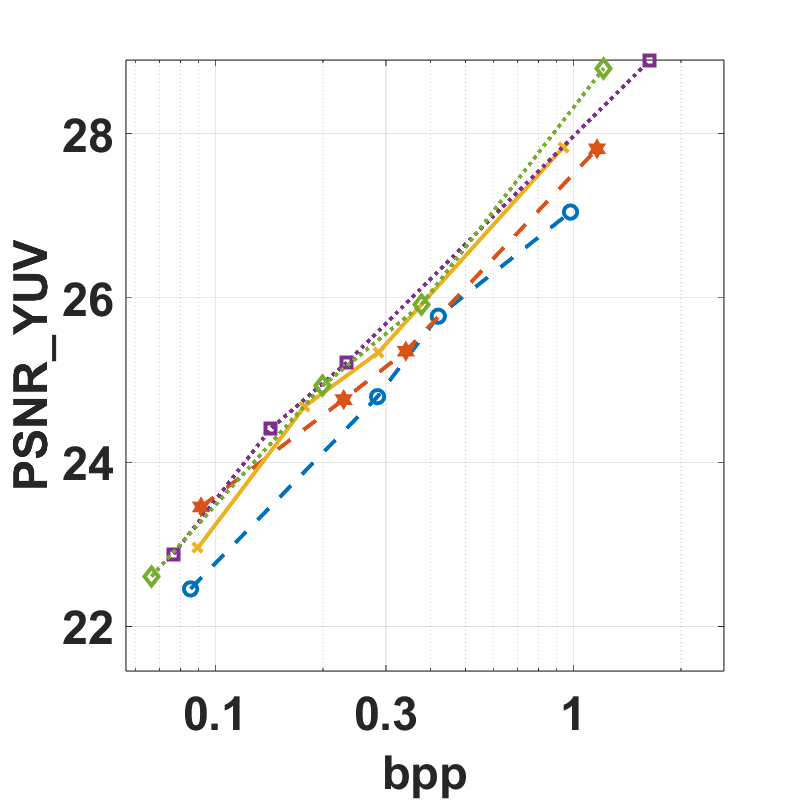}}\hfill
    \subfloat[\textit{GraphSIM (JPEG PCC VM)}]{\includegraphics[width=0.32\linewidth]{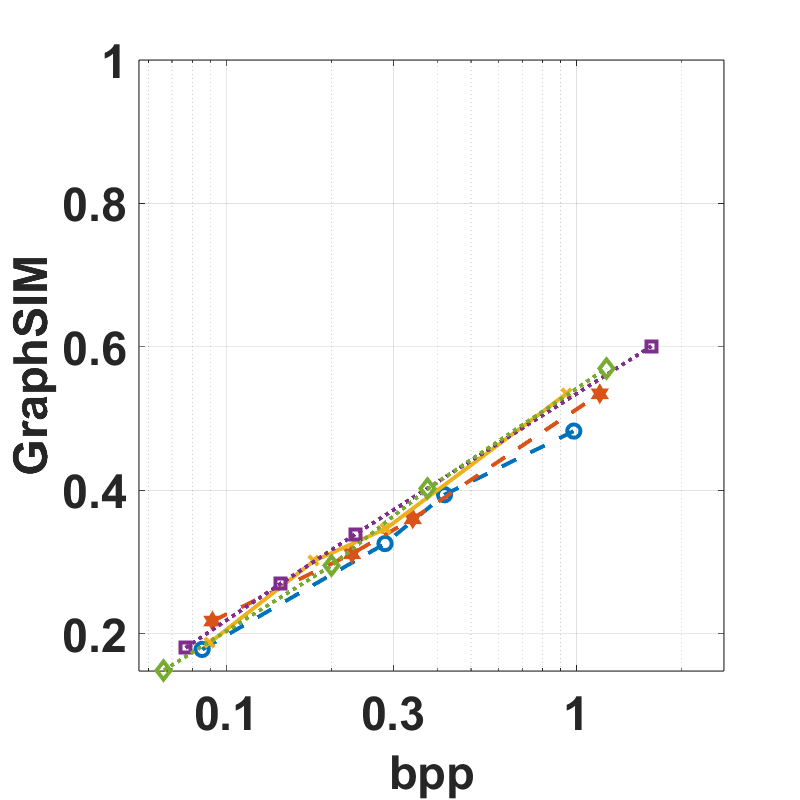}}
    \subfloat[\textit{1 - PCQM (JPEG PCC VM)}]{\includegraphics[width=0.32\linewidth]{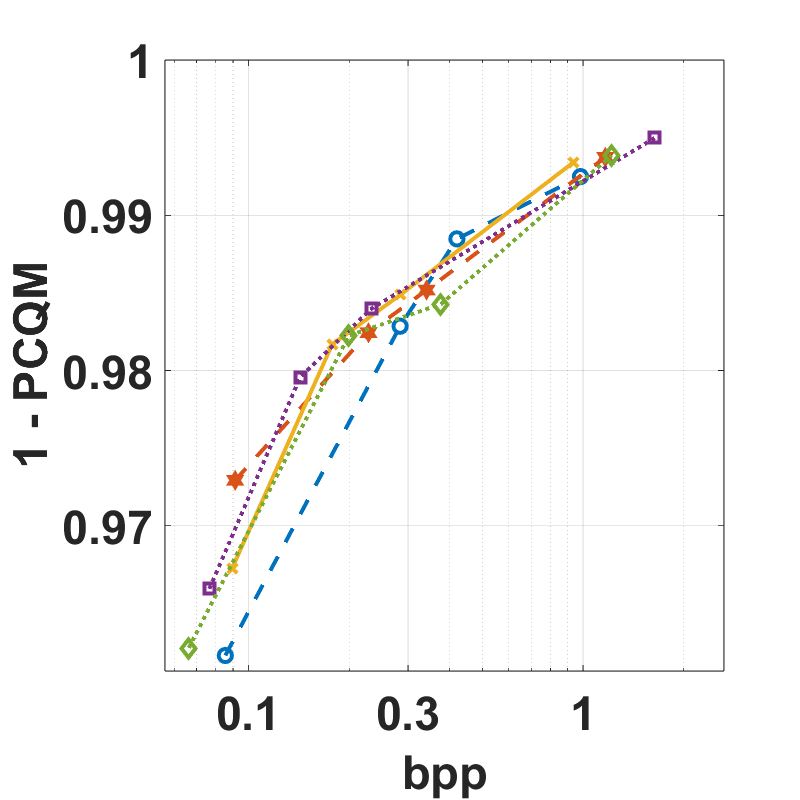}}\hfill\\
    \includegraphics[width=\linewidth]{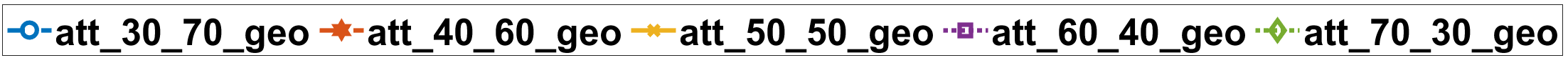}
    \caption{Plots of PSNR YUV, 1 - PCQM and GraphSIM metrics for the \textit{Bouquet} point cloud.}
    \label{fig:Metrics}
        \end{center}
\end{figure}

\subsection{Dataset encoding}\label{sec:dataEncoding}

For this study, the MPEG standard G-PCC~\cite{VPCCandGPCC} and the JPEG Pleno Point Cloud Verification model (JPEG PCC VM)~\cite{VM-CD} were selected. 

\subsubsection{G-PCC}
To encode geometry, G-PCC uses either octree or trisoup. The octree mode was selected, as it is usually the most widely used and outperforms trisoup~\cite{icip2020}. 
In terms of color encoding, G-PCC allows the use of the Regional Adaptive Hierarchical Transform (RAHT)~\cite{Queiroz2016a}, the predicting transform~\cite{VPCCandGPCC}, and the lifting transform~\cite{VPCCandGPCC}.

The primary concept of RAHT is to use the attribute values at a lower octree level to predict the values at the subsequent level. The predicting/lifting transform is a combination of the predicting transform and the lifting transform. The predicting transform uses an interpolation-based hierarchical nearest-neighbor prediction algorithm. It serves as the basis for the definition of the lifting transform, incorporating an additional update and lifting step.

The codec contains two main parameters to control the geometry and color attribute distortion, notably the positionQuantizationScale (pQs) and Quantization Parameter (QP), respectively. Different values were tested in order to achieve the desired tradeoffs. Furthermore, four target bitrates were considered, notably 0.1, 0.3, 1, and 3 bits per point (bpp) $\pm10\%$, representing distinct quality levels.

\subsubsection{JPEG Pleno Point Cloud Verification Model}
The JPEG Pleno Point Cloud Verification Model (JPEG PCC VM)~\cite{VM-CD} is a learning based solution that is able to encode both geometry and attributes.
The codec divides a point cloud into blocks for geometry encoding, which is then transformed into a latent representation using sparse convolutions~\cite{choy20194d}, creating coordinates and feature vectors, which are coded using octree and a variational autoencoder.
For attribute coding, the color attributes are mapped to the decoded point cloud.
The V-PCC model is used to generate point cloud projections~\cite{VPCCandGPCC}. Those projections are then encoded with JPEG-AI~\cite{JPEGAI,CD-JPEGAI}.
The codec allows four different inputs, notably Block Size (BS), which controls the size of the blocks that will be encoded (e.g.64x64x64), Sampling Factor (SF), which controls the scaling of the point cloud coordinates, super resolution (SF), which densifies the point cloud in order to increase the quality, at no rate cost and color Index (CI), which controls the quality of JPEG AI, when encoding the point cloud attributes. 
It is very difficult to control the tradeoff bitrate for the  JPEG PCC VM codec, as it was not designed with that functionality.
Given this, several combinations of the parameters were evaluated, and the ones that resulted in approximately the target tradeoff were selected. 

Visual examples of the \textit{Bouquet} point clouds encoded with G-PCC using the Predicting/Lifting transform and the JPEG PCC VM are shown in Figs \ref{fig:VIS_pred} and \ref{fig:VIS_VM}, respectively. In these figures, the different artifacts created by the different tradeoffs can be observed. For G-PCC, higher geometry bitrate values generate a precise reconstruction of the geometry, but the colors of the point cloud have less definition than those shown in the tradeoffs with higher attribute bitrate.
The visual examples of the JPEG PCC VM show the same behavior, but even with a higher geometry bitrate, the blocking artifacts typical of deep-learning technology are still present.

\begin{table}[!t]
    \centering
    \Huge
    \caption{Average BD-Metrics and BD-Rate for each point cloud, using attribute/geometry tradeoff of 50\%/50\% as a reference for each of the tested codecs.}
   \resizebox{\linewidth}{!}{%
    \begin{tabular}{|c|c|c|c|c|c|c|c|c|c|c|c|}
    \hline
         & \multicolumn{2}{|c|}{\textbf{PSNR YUV}} & \multicolumn{2}{c|}{\textbf{PCQM}} & \multicolumn{2}{c|}{\textbf{GraphSIM}}\\ \hline
        \textbf{\makecell[c]{Attribute\\Geometry \\ Tradeoff}} & \textbf{BD-PSNR} & \textbf{BD-Rate} & \textbf{BD-Metric} & \textbf{BD-Rate}  & \textbf{BD-Metric} & \textbf{BD-Rate}  \\ \hline
        &\multicolumn{6}{c|}{\textbf{G-PCC Predicting/Lifting}} \\ \hline
        70\%/30\% & -0.019 & 1.604\% & -3.347E-04 & 17.838 \% & -3.663E-02 & 34.289\%  \\ \hline
        60\%/40\% & 0.117 & -3.558\% & 5.232E-04 & 4.555 \% & -5.618E-03 & 8.331\%  \\ \hline
        40\%/60\% & -0.325 & 11.525\% & -1.001E-03 & 13.174 \% & -9.095E-03 & 10.013\% \\ \hline
        30\%/70\% & -0.721 & 30.293\% & -2.241E-03 & 24.388 \%  &-1.914E-02 & 17.048\% \\ \hline \hline
        &\multicolumn{6}{c|}{\textbf{G-PCC RAHT}} \\ \hline
        70\%/30\% & -0.097 & 3.746\% & -3.544E-04 & 15.264\% & -0.036 & 24.385\%  \\ \hline
        60\%/40\%& 0.080 & -1.214\% & 2.501E-04 & -1.287\% & -0.016 & 7.131\%  \\ \hline
        40\%/60\% & -0.331 & 12.274\% & -1.426E-03 & 15.932\% & -0.013 & 6.785\% \\ \hline
        30\%/70\% & -0.833 & 26.979\% & -2.911E-03 & 34.669\% & -0.030 & 18.499\% \\ \hline \hline
        &\multicolumn{6}{c|}{\textbf{JPEG PCC VM}} \\ \hline
        70\%/30\% &0.144 & -6.612\% & -8.62E-04 & -23.632\%   & -0.002 & 30.167\%\\ \hline
        60\%/40\% & -0.272 & -12.952\% & -1.36E-03 & -26.361\% & -0.015 & 193.519\%  \\ \hline
        40\%/60\% & -0.485 & 0.661\% & -9.59E-04 & -3.119\% & -0.039 & 56.771\% \\ \hline
        30\%/70\% &-0.820 & 22.567\% & -1.28E-03 & -8.695\% & -0.046 & 56.661\% \\ \hline
    \end{tabular}%
    }\label{tab:BDRates}
\end{table}

\subsection{Objective evaluation}
Prior to the objective evaluation, expert viewing was conducted by three experts to understand which metrics were more accurate for quality representation. Up to the authors knowledge, there is no research work where metrics were validated for distortions in the point cloud color quality evaluation.
Using expert viewing, it was concluded that the best metrics for this specific evaluation are in the following order: PSNR YUV~\cite{PSNRYUV}, GraphSIM~\cite{QiYangGraphSIM2022}, and PCQM~\cite{MeynetPCQM}.

The PSNR YUV is calculated by comparing color values between reference and distorted point clouds using the nearest neighbor algorithm. Each color channel's error is computed based on Euclidean distance, and the overall PSNR YUV is calculated by weighting each channel ($Y:U:V = 6:1:1$).

GraphSIM extracts geometry key points from the reference point cloud, creating neighborhoods (in both the reference and distorted point clouds) around those points. Then, two chroma and one luminance gradient features are extracted~\cite{QiYangGraphSIM2022}.

PCQM~\cite{MeynetPCQM} uses a weighted linear combination of curvature and luminance information to predict the visual quality of a distorted point cloud.


Fig. \ref{fig:Metrics} shows the results obtained by the objective point cloud quality metrics for the \textit{Bouquet} point cloud. The PSNR YUV metric computes the error of each color channel in the YUV color space. However, the number of points that are present in the reconstructed point cloud affects the metric. The plots reveal that when using an attribute/geometry tradeoff of 70\%/30\% using the lifting transform, the metric shows worse results than 60\%/40\%, even though it has a higher attribute bitrate. When using RAHT, a similar behavior can be observed. The geometry/attribute tradeoff of 60\%/40\% usually shows the best results. 
Moreover, the metric evaluates that the tradeoffs of 60\%/40\% and 70\%/30\% have very similar quality for the JPEG PCC VM.

The GraphSIM metric shows almost no change between different tradeoffs when compared with the other metrics.
Upon closer inspection, it can be observed that for both the predicting/lifting and RAHT algorithms, the worst performing tradeoff was the attribute/geometry ratio of 70\%/30\%. Although this metric mainly considers chroma and luminance, a poor geometry reconstruction may have a strong impact on the neighborhood generation, thus leading to lower quality values. Conversely, for the JPEG PCC VM, that specific tradeoff was one of the best performing ones, at higher bitrates.

At higher bitrates, PCQM exhibits better quality for tradeoffs that allocate a higher bitrate for color, although typically there are negligible differences between metric values.
Notably, the metric shows higher quality results for the 30\%/70\% tradeoff at high bitrates, which was not verified for the other metrics.

To further complement the study, the Bjontegaard deltas \cite{Bjontegaard} for the PSNR and bitrate are computed for the tested codecs.
They are computed for each point cloud, and the results are averaged, helping to understand the performance achieved by the different tradeoffs.
The tradeoff of 50\%/50\% is used as a reference. The results for predicting/lifting, RAHT, and JPEG PCC VM are shown in Table \ref{tab:BDRates}.

Analyzing the results for JPEG PCC VM, it may not be clear which tradeoff achieves the best results. The results of PNSR YUV clearly show that the highest gains are achieved by the attribute/geometry tradeoff of 70\%/30\%. 
Both PCQM and GraphSIM show a similar performance for the two tradeoffs of 60\%/40\% and 70\%/30\%. 
Considering the performed expert viewing session, that considered PSNR YUV as the most reliable metric of the three, would lead to the conclusion that the best tradeoff is 70\%/30\%. However, this result needs further analysis, and possibly can only be concluded with a subjective evaluation.


\section{Conclusions}
This paper presents an objective quality assessment of the attributes and geometry coding bitrate tradeoffs for G-PCC and the JPEG PCC VM when encoding colored static point clouds.
The reported results indicate that better quality is obtained when a benefit is given to the coding bitrate of the color attributes over the geometry information.
Nevertheless, the performance gains are very small.
The improvements in quality were also observed with expert viewing.
For future work, a subjective quality evaluation is planned to perceptually assess how these tradeoffs affect the compression performance and to evaluate how accurately the metrics represent the subjective quality scores.

\bibliographystyle{IEEEtran}
\bibliography{refs}
\end{document}